\documentstyle[aps,twocolumn,prl,epsf]{revtex}

\begin{document}

\newcommand{\be}{\begin{equation}}
\newcommand{\ee}{\end{equation}}
\newcommand{\bn}{\begin{eqnarray}}
\newcommand{\en}{\end{eqnarray}}

\draft

\twocolumn[\hsize\textwidth\columnwidth\hsize\csname @twocolumnfalse\endcsname

\title{Insulator-Metal transition in the Doped 
$3d^{1}$ Transition Metal Oxide $LaTiO_{3}$}

\author{L. Craco$^1$, M. S. Laad$^1$, S. Leoni$^2$, and
E. M\"uller-Hartmann$^1$}

\address{$^1$Institut f\"ur Theoretische Physik, Universit\"at zu K\"oln, 
Z\"ulpicher Strasse, 50937 K\"oln, Germany  \\
$^2$Max-Planck-Institut f\"ur Chemische Physik fester Stoffe, 01187 Dresden, 
Germany}

\date{\today}
\maketitle

\widetext

\begin{abstract}
The doping induced insulator-metal transition in $La_{1-x}Sr_{x}TiO_{3}$ is 
studied using the ab-initio LDA+DMFT method.  Combining the LDA bandstructure 
for the actual, distorted structure found recently with multi-orbital
DMFT to treat electronic correlations, we find: $(i)$ ferro-orbital 
order in the Mott insulating state without orbital degeneracy, $(ii)$ a 
continuous filling induced transition to the paramagnetic metal (PM) with 
$x$, and $(iii)$ excellent quantitative agreement with published photoemission 
data for the case of $6\%$ doping.  Our results imply that this system can be 
described as a Mott-Hubbard system without orbital (liquid) degeneracy. 
\end{abstract}
\pacs{PACS numbers: 
71.27.+a, 
71.15.Mb,  
79.60.-i  
}

]

\narrowtext

The early TMO $LaTiO_{3}$ is an interesting case of a system exhibiting 
filling-driven insulator-metal transition. Long thought of as a text-book 
example of a Mott insulator with antiferromagnetic order, current interest 
in orbital correlations in TMOs has reopened the subject of the physics of 
$LaTiO_{3}$~\cite{[1]}. 

  According to~\cite{[2]}
this material with $Ti^{3+}(3d^{1})$ configuration 
is an electron analog of the cuprates, with $G$-type AF spin order below 
$T_{N} \simeq 150~K$, and with a reduced spin moment in the AF 
state.  Recent neutron scattering measurements reveal an {\it isotropic}
magnon dispersion, with a small spin gap, $\Delta_{s}=3.3~meV$, putting 
strong constraints on an acceptable mechanism for the observed 
antiferromagnetism~\cite{[2]}.
The fact that the single electron in $LaTiO_{3}$ resides in the $t_{2g}$ 
orbitals (less prone to Jahn-Teller distortions) suggested that 
the orbital moment might not be fully quenched in this case.  Attempts to 
explain the strongly reduced spin moment of $\mu=0.46 \mu_{B}$~\cite{[2]}
assuming an unquenched orbital moment aligned antiparallel to the spin moment 
(spin-orbit coupling in the $t_{2g}$ shell of $Ti^{3+}$) led to a spin gap 
much larger than that of $3.3~meV$ observed in INS~\cite{[2]},  
excluding an orbital contribution to the ordered moment.  
Moreover, ferro-orbital order (FOO) has recently been suggested~\cite{[3]}, 
in complete accord with the $G$-type AF spin order below $T_{N}$, implying
lifting of the $t_{2g}$ orbital degeneracy.

The metallic phase in $La_{1-x}Sr_{x}TiO_{3}$~\cite{[1]}
has been shown to be a canonical strongly correlated Fermi liquid~\cite{[1]},  
putting additional constraints on an acceptable model for this system.  In 
particular, this implies a non-degenerate ground state in the paramagnetic 
metal (PM) phase, since any remnant of orbital degeneracy
(if it was hypothesized to exist) would lead to violation of Landau's Fermi 
liquid hypothesis, and lead to a non-FL metallic state, in contradiction with
observations.  AF LRO is rapidly suppressed upon $Sr$ doping, and 
classic signatures of the Brinkman-Rice (BR) correlated FL are seen close to 
the MIT, suggesting that the AFI state is a Mott-Hubbard insulator and that the
AFI-PM transition is a Mott-Hubbard transition. Given the incomplete 
understanding of the AFI, the mechanism of the MIT still constitutes an open 
problem for theory, in spite of much recent activity~\cite{[4]}.

  Two alternatives were proposed to account for the puzzling observations 
in the AFI phase.  Mochizuki {\it et al.}~\cite{[5]} presented a  
model for the AF in $LaTiO_{3}$ using a $GdFeO_{3}$ distortion, which, 
however, was not observed.  Khaliullin {\it et al}~\cite{[6]} proposed a 
novel theoretical idea based on an orbital liquid, but the short-ranged 
orbital RVB-like fluctuations were not observed either~\cite{[3]}.  Moreover, 
it has been shown recently that long range AF order is theoretically 
{\it incompatible} with an orbital liquid state~\cite{[7]}, deepening the 
controversy regarding the origin of $G$-type AF order in $LaTiO_{3}$.

Here, following new work~\cite{[8]} pinpointing the real crystal structure 
(RCS) of $LaTiO_{3}$, we propose a detailed theoretical scenario unifying 
the various experimentally observed features in the insulating (I) as well 
as the correlated metallic (M) states.  Employing the state-of-the-art 
technique LDA+DMFT (IPT) (combining the real bandstructure 
of $LaTiO_{3}$ with dynamical correlation effects of local 
Coulomb interactions), we show how a consistent description 
of the doping-induced insulator-metal (I-M) transition, as well as very good 
quantitative agreement with photoemission results is obtained.
 
The RCS of $LaTiO_{3}$ (i.e, $Pbnm$ space group) involves a structural 
distortion arising from the tilting of the $TiO_{6}$ octahedra around the 
$[110]_{c}$ axis followed by their rotation around $c$, beginning from the 
cubic perovskite structure.  The specific distortion of the $TiO_{6}$ 
octahedra in $LaTiO_{3}$ is characterized by a large splitting in the $O2-O2$ 
edge lengths of the octahedral basal plane, giving rise to elongation along 
$a$ and shrinking along $b$.  This implies FOO and a lifting of $t_{2g}$ 
orbitaldegeneracy.  Starting from a local picture in the 
$d^{1}$ configuration, the occupied orbital is constructed as: 
$|a\rangle=0.77(\frac{|xz+yz\rangle}{\sqrt{2}}) \pm 0.636|xy\rangle$, which 
is non-degenerate~\cite{[8]}.  We have carried out a local density 
approximation (LDA) study based on the RCS described above.

\begin{figure}[htb]
\epsfxsize=3.6in
\epsffile{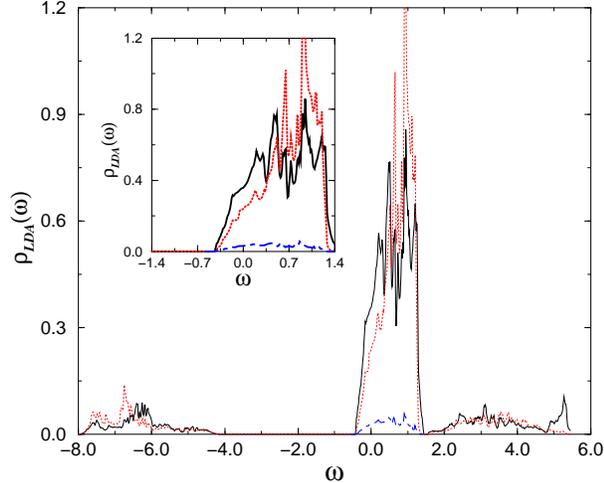}
\caption{LDA partial densities of states (DOS) for the real crystal structure
of $LaTiO_{3}$. Full line denotes the DOS for the non-degenerate 
ground-state orbital, the dotted line denotes the DOS for the higher (almost 
two-fold degenerate) lying $t_{2g}$ orbitals.  The dot-dashed line denotes 
the small $e_{g}$-DOS around $E_{F}$.  Higher-lying $e_{g}$ and $O$-2p bands 
are not included in the Figure.}
\label{fig2}
\end{figure}

To obtain the band DOS, the one-electron scalar relativistic 
Schr\"odinger equation was solved self-consistently within the 
LDA, using the linear muffin-tin orbitals (LMTO) scheme in 
the atomic sphere approximation, with combined correction terms~\cite{sl1}. 
The exchange-correlation potential was parametrized according to the von 
Barth-Hedin scheme~\cite{sl3,sl4}. Self-consistency was reached on performing 
calculations on a {\bf k}-mesh of dimension $16 \times 16 \times 16$. 
The radii of the atoms in the real RCS were chosen to minimize their overlap. 
Values of $r=3.7~(La)$, $r=2.12~(O)$ and $r=2.75~(Ti)$ a.u. 
were found to be useful.

The total LDA density of states, 
$\rho(\omega)=\sum_{{\bf k},\alpha}\delta(\omega-\epsilon_{\alpha}({\bf k}))$, 
with the LDA (one-electron) Hamiltonian given by

\bn
\nonumber
H_{0}=\sum_{{\bf k},\alpha}\epsilon_{\alpha}({\bf k})
c_{{\bf k}\alpha}^{\dag}c_{{\bf k}\alpha} \;.
\en
 
Several interesting points are apparent from the LDA results. First, the 
$t_{2g}$ splitting is {\it not} negligible, and the character of 
the occupied ground state orbital $|\alpha\rangle$ (not shown) is in accord 
with the result of Ref.~\cite{[8]}.  The LDA bandwidth of the $t_{2g}$ bands 
is about $2~eV$, and the charge transfer energy (LDA) is large, 
$\Delta_{LDA} \simeq 7.0~eV$, while the onsite Hubbard $U \approx 6.0~eV$, 
putting $LaTiO_{3}$ in the Mott-Hubbard category, as expected~\cite{[1]}.  
Further, the Hund coupling, $J_{H}\simeq 1~eV$ (relevant in the excited 
$d^{2}$ configuration, with the two $d$ electrons on different orbitals, 
realized upon $d-d$ hopping), and the inter-orbital coupling, 
$U'=(U-2J_{H})\simeq 4.0~eV$.  These results (see Fig.~\ref{fig2})
do not represent an insulating state, this requiring a proper treatment of 
correlation effects in the RCS.  Given the FOO found in $LaTiO_{3}$, a 
description in terms of a single-band model for the $|\alpha\rangle$ band is 
untenable, and requires a full multi-orbital dynamical mean field calculation. 
To avoid double counting of interactions already treated on the average by 
LDA, we follow~\cite{[4]} to write,

\bn
\nonumber
H_{0}'=H_{0} + \sum_{i\alpha\sigma}\epsilon_{i\alpha\sigma}^{0}
n_{i\alpha\sigma} \;,
\en
with $\epsilon_{i\alpha\sigma}^{0}=\epsilon_{i\alpha}-
U(n_{\alpha -\sigma}-\frac{1}{2}) + \frac{J_H}{2}(n_{\alpha\sigma}-1)$.

The full many-body Hamiltonian is now given by,

\bn
\nonumber
H &=& H_{0}^{'} + U\sum_{i,\alpha} n_{i\alpha\uparrow} n_{i\alpha\downarrow} + 
U'\sum_{i\alpha\beta} n_{i\alpha} n_{i\beta} \\ \nonumber
&-& J_{H}\sum_{i\alpha\beta} {\bf S}_{i\alpha} \cdot {\bf S}_{i\beta} \;.
\en

We solve this multiband model in $d=\infty$ using multi-orbital iterated 
perturbation theory (IPT)~\cite{[LMH]}.  Assuming no symmetry breaking in 
the spin/orbital sectors, we have $G_{\alpha\beta\sigma\sigma'}(\omega)=
\delta_{\alpha\beta}\delta_{\sigma\sigma'}G_{\alpha\sigma}(\omega)$ and 
$\Sigma_{\alpha\beta\sigma\sigma'}(\omega)=\delta_{\alpha\beta}
\delta_{\sigma\sigma'}\Sigma_{\alpha\sigma}(\omega)$. Further, from the 
LDA results, the DOS for the $e_{g}$ bands is small near $E_{F}$.  Taking 
this together with the fact that the $O-2p$ bands lie $6~eV$ away from 
$E_{F}$, we keep only the $t_{2g}$ bands in what follows.

The DMFT solution involves~\cite{[AG]} $(i)$ replacing the lattice model by a 
self-consistently embedded multi-orbital, asymmetric Anderson impurity model, 
and, $(ii)$ the selfconsistency condition requiring the local impurity 
Green function to be equal to the local GF for the lattice. In $LaTiO_{3}$, 
in the $3d^{2}$ configuration reached by hopping, $U', J_{H}$ scatter 
electrons between different $t_{2g}$ orbitals, so only the {\it total} 
number, $n_{t_{2g}}^d=\sum_{\alpha}n_{t_{2g},\alpha}^d$ is
conserved in a way consistent with Luttinger's theorem.
The calculation follows the philosophy of the one-orbital IPT, with the GFs 
and the self-energies being matrices in the orbital indices.  
The full set of equations for the multi-orbital case are the same as those
appearing elsewhere~\cite{[Held]}, and we do not reproduce them here.
Local self-energies computed from this generalized IPT formulation 
exactly satisfy the Friedel-Luttinger constraint for arbitrary 
band-fillings, guaranteeing the correct low-energy form of the spectral
function. The correct form of the spectrum at high energy is ensured by 
the (exact) selfconsistent determination of the first few moments of the 
spectral function.  These equations are solved selfconsistently with the LDA
DOS as input, until convergence is achieved. 

\begin{figure}[htb]
\epsfxsize=3.6in
\epsffile{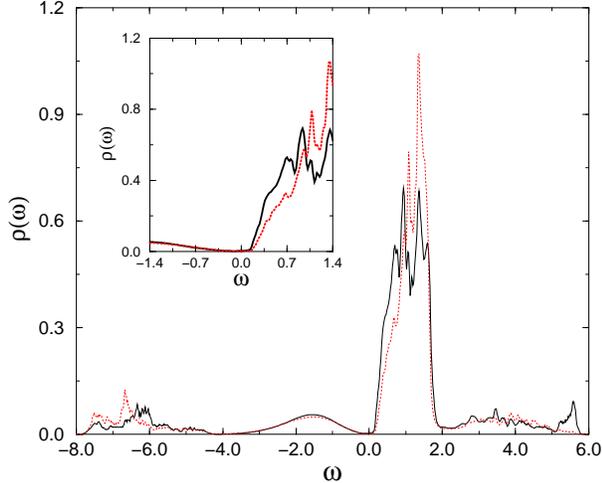}
\caption{LDA+DMFT results for the partial DOS in the insulating state, 
obtained using $U=6.0~eV$, $J_{H}=1.0~eV$, and $U'=4.0~eV$ in the real 
bandstructure of $LaTiO_{3}$.}
\label{fig3}
\end{figure}

We now present our results. $\;$ Figure~\ref{fig3} shows the total spectral 
function for the paramagnetic insulating (PI) state of $LaTiO_{3}$.   This 
does not correspond to the true ground state, which is a $G$-type AFI, 
but to the PI (Mott) state above $T_{N}$. $\;$ The Mott-Hubbard gap is read  
off directly from the DOS as $\Delta_{G}^{I} \simeq 0.4~eV$.  $\;$
Further, the renormalized value of the $t_{2g}$ splitting is computed to 
\hspace{-0.3cm}be $0.15~eV$, implying a reduction in the charge transfer 
energy, $\Delta_{ct}=(\epsilon_{p}-\epsilon_{d})$ from $7.0~eV$ in the LDA to 
$6.64~eV$ in the correlated solution.  This is likely to be an overestimate, 
since further downward renormalization of $\Delta_{ct}$ will result 
from additional Hartree shifts arising from inclusion of Madelung ($p-d$ 
interaction) terms.  These results constitute a concrete resolution of the 
questions posed recently~\cite{[9]}, namely, the mechanism for the 
generation of a small-gap Mott-Hubbard insulator in a situation where
the Coulomb interactions are comparable to the LDA bandwidth.  In view of our 
results, dynamical effects of coupled spin/orbital correlations in the real
structure of $LaTiO_{3}$ are of crucial importance for a consistent resolution
of this issue.  We also find the occupation of the lowest $t_{2g}$ orbital,
$n_{|\alpha\rangle}=0.17$, slightly higher than that of the other two 
higher-lying (by $0.15~eV$) orbitals in the $t_{2g}$ sector.

These results are consistent with having ferro-orbital order (FOO) in 
$LaTiO_{3}$~\cite{[3]}. Further, since the ground 
state occupied orbital, $|\alpha\rangle=a_{1}|xz+yz\rangle + a_{2}|xy\rangle$ 
(consistent with Ref.~\cite{[8]}) is {\it not} oriented along one of the 
bonds, the resulting AF exchange is not expected to be very anisotropic.  A 
detailed calculation of the Heisenberg superexchange based on the LDA+DMFT 
results, along with proper incorporation of spin anisotropies, is in progress 
and will be reported elsewhere.
 
To model the paramagnetic metallic (PM) state, reached (continuously from the 
PI in our approach) upon doping $LaTiO_{3}$
with $Sr$, we extended our calculation for the PI to: $(i)$ incorporate 
the change in band-filling along the same lines as for generalized IPT for the 
single band case.  This implies computing the chemical potential shift, as 
required by Luttinger's theorem, and, $(ii)$ considering the effects of $Sr$
doping induced static disorder (the difference of the on-site potentials on 
$Sr$ substitution is $\Delta v \simeq 4.0~eV$~\cite{[10]}) 
using the coherent-potential approximation (CPA) combined 
selfconsistently~\cite{[11]} with the multi-orbital IPT described above.  
Furthermore, to obtain quantitative consistency with spectroscopic results, 
we found it necessary to include the (small near $E_{F}$) DOS of the $e_{g}$ 
bands found in the LDA (Fig.~\ref{fig2}), as well (see below).

Generically, hole doping the (Mott) PI creates a very 
sharp, quasicoherent Fermi liquid peak at the renormalized $E_{F}$, with 
dynamical spectral weight transfer over a large energy scale from high to low
energy (caused by $U, U', J_{H}$ within DMFT), driving a first-order Mott 
transition in the generic case.  In this situation, static, potential 
disorder is expected to lead to a moderate broadening of the sharp FL peak, 
with spectral weight transfer from low to high energy~\cite{[11]}, and, more 
importantly, to convert the first-order Mott transition into a continuous one.
Thus, the character of the MIT, as well as details of the actual one-particle 
spectrum depend sensitively on how the competition between Mott-Hubbard 
localization and itinerance in a multi-band system is affected by doping 
induced static disorder.

\begin{figure}[htb]
\epsfxsize=3.6in
\epsffile{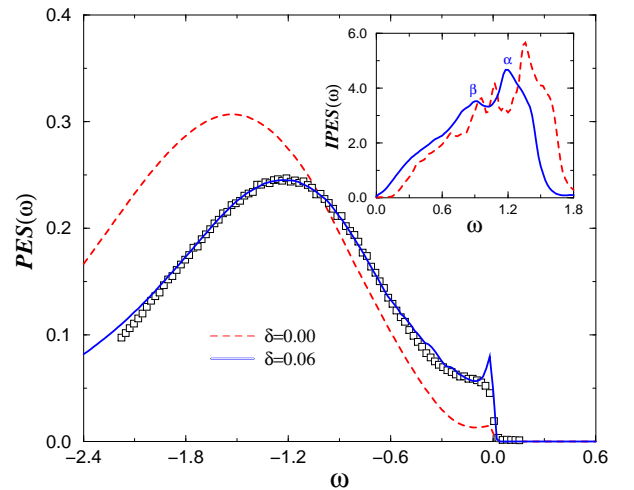}
\caption{The photoemission (PES) and inverse photoemission (IPES: in inset) 
spectra for both, pure (dashed) and doped (solid) $LaTiO_{3}$.  Excellent 
quantitative agreement with experimental results (PES: squares) of 
Ref.~[18] is clear.}
\label{fig4}
\end{figure}

With this in mind, we show the photoemission (PES) and inverse-PES (IPES) 
spectra for the lightly ($\delta \equiv x+y=0.06$) doped, PM 
state of $La_x Sr_{1-x}TiO_{3+y/2}$~\cite{[12]}, along with the 
experimental result for $I_{PES}(\omega)$ in Fig.~\ref{fig4}. Clearly, 
excellent quantitative 
agreement between theory and experiment is found over the entire range 
$-2 \le \omega \le E_{F}$.  $\;$ In particular, the lower Hubbard band, the 
quasicoherent FL feature, as well as the details of the lineshape are all 
in excellent agreement with experiment. $\;$ In the IPES spectrum, we are 
able to resolve the $\alpha, \beta$ peaks 
\hspace{-0.3cm}alongwith the much greater signal 
intensity and the correct detailed form of the lineshape~\cite{[13]}.
Thus, our results represent a {\it quantitatively} accurate representation 
of the complete one-electron spectral function in the PM phase of 
$La_{1-x}Sr_{x}TiO_{3+y/2}$.  Lastly, from the Fermi liquid self-energies
(not shown), the effective mass renormalization is estimated to be 
$m^{*}/m_{b}=1.5$ for $\delta=0.06$ in good quantitative agreement with the 
frequency-dependent renormalization estimated from PES~\cite{[12]}. Actually, 
the total renormalization includes additional renormalization (factor of $2$)
coming from the ${\bf k}$-dependence of the self-energies.  On general 
grounds, we do expect such renormalization in a case of proximity to the AF 
transition; theoretically, an extension to include non-local correlations is 
required to address this issue. We have not done this here.

We stress essential differences between our work and earlier works.  In 
Ref.~\cite{[4],[5]}, an incorrect crystal structure was used for the LDA 
calculation. Moreover, the strong $J_{H} \simeq 1.0~eV$ along with the 
inter-orbital $U'$ seem to have been completely ignored in their correlation
(DMFT) calculations. Turning to Ref.~\cite{[6]}, we believe, given the fact 
that a multitude of experiments reveal correlated (Brinkman-Rice) FL 
responses (PM), that the basic Landau hypotheses force one to abandon a 
picture based on orbital degeneracy in the PM phase as well.  We re-emphasize 
that AF-LRO has recently been shown to be impossible~\cite{[7]} for a 
$3d^{1}$ system with triple orbital ($t_{2g}$) degeneracy.  Very recently, 
Pavarini {\it et al.}~\cite{[21]} have studied a host of $3d^{1}$ systems 
within LDA+DMFT, obtaining similar conclusions for the Mott insulating state
in $LaTiO_{3}$.  However, the $Sr$-doping induced I-M transition has not been 
studied there.  Additionally, using QMC to solve
the DMFT equations forces a restriction to high-$T=770~K$~\cite{[21]}.  Given 
the strong $T$ dependence of the correlated Fermi liquid scale within DMFT, as
well as the much lower FL coherence scale observed experimentally in doped 
$LaTiO_{3}$~\cite{[1],[AG]}, (in contrast to QMC results), we believe that a 
quantitative comparison with experiment at low $T$ requires the solution of the
DMFT equations at lower $T \simeq 150~K$, as we have done.
Finally, we draw attention to the excellent 
quantitative agreement between theory and experiment for the one-electron
spectral function, obtained using LDA+ (multi-orbital) DMFT.  To our knowledge,
this is the first calculation achieving such unprecedented agreement between
experiment and theory for this system.
 
To conclude, we have studied the doping induced Mott transition in the
$3d^{1}$ system $La_{1-x}Sr_{x}TiO_{3}$ using the ab-initio LDA+DMFT method.  
Starting from the real crystal structure (with lifting of $t_{2g}$ orbital 
degeneracy), we have explicitly demonstrated the nature of the carrier 
driven MIT in this system.  Our results support ferro-orbital order  
in the insulating state, consistent with recent suggestions~\cite{[3]}, and 
with the observed $G$-type AF order. In conjunction with [8, 21], these results 
favor a description of the physics of $LaTiO_{3}$ in a picture of 
non-degenerate $t_{2g}$ orbitals.   
Within this picture, in the PM state, excellent quantitative 
agreement with published PES {\it and} IPES results is obtained, providing a 
complete description of the evolution of one-particle excitations across 
the MIT.  Similar ideas can be fruitfully applied to study the doping-driven 
MI transitions in other TMO-based systems~\cite{[20]} with coupled
spin-orbital correlations.

\acknowledgements
MSL and LC acknowledge fruitful discussions with D. I. Khomskii in course of 
this work.
SL wishes to aknowledge A. Yaresko for his help with the LMTO program and
for valuable discussions. 
Work carried out under the auspices of the Sonderforschungsbereich 608 of the
DFG (K\"oln), and the MPI-CPfS (Dresden).

\end{document}